\documentclass[aps,prl,twocolumn,superscriptaddress,longbibliography,floatfix]{revtex4-2}
\usepackage{graphicx}
\usepackage{amsmath,amssymb}
\usepackage[T1]{fontenc}
\usepackage[utf8]{inputenc}
\usepackage{newtxtext,newtxmath}
\usepackage{newunicodechar}
\newunicodechar{−}{\ensuremath{-}}
\newunicodechar{°}{\ensuremath{^{\circ}}}
\newunicodechar{±}{\ensuremath{\pm}}
\newunicodechar{μ}{\ensuremath{\mu}}
\newunicodechar{×}{\ensuremath{\times}}
\newunicodechar{λ}{\ensuremath{\lambda}}
\usepackage{url}
\usepackage{hyperref}
\hypersetup{colorlinks=true,allcolors=blue}

\begin{document}
\title{Tidal potential measured with distant optical clocks}
\author{Lei Hou}\thanks{These authors contributed equally to this work.}\affiliation{Hefei National Laboratory, University of Science and Technology of China, Hefei, China}\affiliation{Shanghai Research Center for Quantum Sciences and CAS Center for Excellence in Quantum Information and Quantum Physics, University of Science and Technology of China, Shanghai, China}
\author{Xiang Zhang}\thanks{These authors contributed equally to this work.}\affiliation{Hefei National Laboratory, University of Science and Technology of China, Hefei, China}\affiliation{Shanghai Research Center for Quantum Sciences and CAS Center for Excellence in Quantum Information and Quantum Physics, University of Science and Technology of China, Shanghai, China}\affiliation{Hefei National Research Center for Physical Sciences at the Microscale and School of Physical Sciences, University of Science and Technology of China, Hefei, China}
\author{Qiang Zhu}\thanks{These authors contributed equally to this work.}\affiliation{Innovation Academy for Precision Measurement Science and Technology, Chinese Academy of Sciences, Wuhan, China}
\author{Xiang-Pei Liu}\thanks{These authors contributed equally to this work.}\affiliation{Hefei National Laboratory, University of Science and Technology of China, Hefei, China}\affiliation{Hefei National Research Center for Physical Sciences at the Microscale and School of Physical Sciences, University of Science and Technology of China, Hefei, China}
\author{Fa-Xi Chen}\thanks{These authors contributed equally to this work.}\affiliation{Hefei National Laboratory, University of Science and Technology of China, Hefei, China}\affiliation{Jinan Institute of Quantum Technology and CAS Center for Excellence in Quantum Information and Quantum Physics, University of Science and Technology of China, Jinan, China}
\author{Wen-Xuan Song}\thanks{These authors contributed equally to this work.}\affiliation{Hefei National Laboratory, University of Science and Technology of China, Hefei, China}\affiliation{Shanghai Research Center for Quantum Sciences and CAS Center for Excellence in Quantum Information and Quantum Physics, University of Science and Technology of China, Shanghai, China}\affiliation{Hefei National Research Center for Physical Sciences at the Microscale and School of Physical Sciences, University of Science and Technology of China, Hefei, China}
\author{Zhuan-Xian Xiong}\affiliation{Hefei National Laboratory, University of Science and Technology of China, Hefei, China}\affiliation{Innovation Academy for Precision Measurement Science and Technology, Chinese Academy of Sciences, Wuhan, China}
\author{Jian-Yu Guan}\affiliation{Hefei National Laboratory, University of Science and Technology of China, Hefei, China}\affiliation{Shanghai Research Center for Quantum Sciences and CAS Center for Excellence in Quantum Information and Quantum Physics, University of Science and Technology of China, Shanghai, China}
\author{Yu-Chen Zhang}\affiliation{Hefei National Laboratory, University of Science and Technology of China, Hefei, China}\affiliation{Shanghai Research Center for Quantum Sciences and CAS Center for Excellence in Quantum Information and Quantum Physics, University of Science and Technology of China, Shanghai, China}\affiliation{Hefei National Research Center for Physical Sciences at the Microscale and School of Physical Sciences, University of Science and Technology of China, Hefei, China}
\author{De-Quan Kong}\affiliation{Hefei National Laboratory, University of Science and Technology of China, Hefei, China}\affiliation{Shanghai Research Center for Quantum Sciences and CAS Center for Excellence in Quantum Information and Quantum Physics, University of Science and Technology of China, Shanghai, China}\affiliation{Hefei National Research Center for Physical Sciences at the Microscale and School of Physical Sciences, University of Science and Technology of China, Hefei, China}
\author{Jia-Hao Zhang}\affiliation{Hefei National Laboratory, University of Science and Technology of China, Hefei, China}\affiliation{Shanghai Research Center for Quantum Sciences and CAS Center for Excellence in Quantum Information and Quantum Physics, University of Science and Technology of China, Shanghai, China}\affiliation{Hefei National Research Center for Physical Sciences at the Microscale and School of Physical Sciences, University of Science and Technology of China, Hefei, China}
\author{Long-Fei Liao}\affiliation{Hefei National Laboratory, University of Science and Technology of China, Hefei, China}\affiliation{Shanghai Research Center for Quantum Sciences and CAS Center for Excellence in Quantum Information and Quantum Physics, University of Science and Technology of China, Shanghai, China}\affiliation{Hefei National Research Center for Physical Sciences at the Microscale and School of Physical Sciences, University of Science and Technology of China, Hefei, China}
\author{Shao-Mao Wang}\affiliation{Hefei National Laboratory, University of Science and Technology of China, Hefei, China}\affiliation{Shanghai Research Center for Quantum Sciences and CAS Center for Excellence in Quantum Information and Quantum Physics, University of Science and Technology of China, Shanghai, China}
\author{Bing Wang}\affiliation{Innovation Academy for Precision Measurement Science and Technology, Chinese Academy of Sciences, Wuhan, China}
\author{De-Zhi Xiong}\affiliation{Hefei National Laboratory, University of Science and Technology of China, Hefei, China}\affiliation{Innovation Academy for Precision Measurement Science and Technology, Chinese Academy of Sciences, Wuhan, China}
\author{Xing-Yang Cui}\affiliation{Hefei National Laboratory, University of Science and Technology of China, Hefei, China}\affiliation{Shanghai Research Center for Quantum Sciences and CAS Center for Excellence in Quantum Information and Quantum Physics, University of Science and Technology of China, Shanghai, China}
\author{Ping Xu}\affiliation{Hefei National Laboratory, University of Science and Technology of China, Hefei, China}\affiliation{Shanghai Research Center for Quantum Sciences and CAS Center for Excellence in Quantum Information and Quantum Physics, University of Science and Technology of China, Shanghai, China}\affiliation{Hefei National Research Center for Physical Sciences at the Microscale and School of Physical Sciences, University of Science and Technology of China, Hefei, China}
\author{Peng Liu}\affiliation{Hefei National Laboratory, University of Science and Technology of China, Hefei, China}\affiliation{Shanghai Research Center for Quantum Sciences and CAS Center for Excellence in Quantum Information and Quantum Physics, University of Science and Technology of China, Shanghai, China}
\author{Jie Li}\affiliation{Shanghai Research Center for Quantum Sciences and CAS Center for Excellence in Quantum Information and Quantum Physics, University of Science and Technology of China, Shanghai, China}\affiliation{Hefei National Research Center for Physical Sciences at the Microscale and School of Physical Sciences, University of Science and Technology of China, Hefei, China}
\author{Xiao Jiang}\affiliation{Hefei National Laboratory, University of Science and Technology of China, Hefei, China}\affiliation{Shanghai Research Center for Quantum Sciences and CAS Center for Excellence in Quantum Information and Quantum Physics, University of Science and Technology of China, Shanghai, China}\affiliation{Hefei National Research Center for Physical Sciences at the Microscale and School of Physical Sciences, University of Science and Technology of China, Hefei, China}
\author{Hua-Qing Zhang}\affiliation{Hefei National Laboratory, University of Science and Technology of China, Hefei, China}\affiliation{Jinan Institute of Quantum Technology and CAS Center for Excellence in Quantum Information and Quantum Physics, University of Science and Technology of China, Jinan, China}
\author{Ling-Xiang He}\affiliation{Hefei National Laboratory, University of Science and Technology of China, Hefei, China}\affiliation{Innovation Academy for Precision Measurement Science and Technology, Chinese Academy of Sciences, Wuhan, China}
\author{Li-Bo Li}\affiliation{Jinan Institute of Quantum Technology and CAS Center for Excellence in Quantum Information and Quantum Physics, University of Science and Technology of China, Jinan, China}
\author{Qi Shen}\affiliation{Hefei National Laboratory, University of Science and Technology of China, Hefei, China}\affiliation{Shanghai Research Center for Quantum Sciences and CAS Center for Excellence in Quantum Information and Quantum Physics, University of Science and Technology of China, Shanghai, China}\affiliation{Hefei National Research Center for Physical Sciences at the Microscale and School of Physical Sciences, University of Science and Technology of China, Hefei, China}
\author{Youchao Xie}\affiliation{Time and Frequency Geodesy Center (TFGC), School of Geodesy and Geomatics, Wuhan University, Wuhan, China}
\author{Lei Wang}\affiliation{Time and Frequency Geodesy Center (TFGC), School of Geodesy and Geomatics, Wuhan University, Wuhan, China}
\author{Sheng-Kai Liao}\affiliation{Hefei National Laboratory, University of Science and Technology of China, Hefei, China}\affiliation{Shanghai Research Center for Quantum Sciences and CAS Center for Excellence in Quantum Information and Quantum Physics, University of Science and Technology of China, Shanghai, China}\affiliation{Hefei National Research Center for Physical Sciences at the Microscale and School of Physical Sciences, University of Science and Technology of China, Hefei, China}
\author{Juan Yin}\affiliation{Hefei National Laboratory, University of Science and Technology of China, Hefei, China}\affiliation{Shanghai Research Center for Quantum Sciences and CAS Center for Excellence in Quantum Information and Quantum Physics, University of Science and Technology of China, Shanghai, China}\affiliation{Hefei National Research Center for Physical Sciences at the Microscale and School of Physical Sciences, University of Science and Technology of China, Hefei, China}
\author{Cheng-Zhi Peng}\affiliation{Hefei National Laboratory, University of Science and Technology of China, Hefei, China}\affiliation{Shanghai Research Center for Quantum Sciences and CAS Center for Excellence in Quantum Information and Quantum Physics, University of Science and Technology of China, Shanghai, China}\affiliation{Hefei National Research Center for Physical Sciences at the Microscale and School of Physical Sciences, University of Science and Technology of China, Hefei, China}
\author{Wen-Bin Shen}\affiliation{Time and Frequency Geodesy Center (TFGC), School of Geodesy and Geomatics, Wuhan University, Wuhan, China}
\author{Baolong Lyu}\affiliation{Hefei National Laboratory, University of Science and Technology of China, Hefei, China}\affiliation{Innovation Academy for Precision Measurement Science and Technology, Chinese Academy of Sciences, Wuhan, China}
\author{Han-Ning Dai}\affiliation{Hefei National Laboratory, University of Science and Technology of China, Hefei, China}\affiliation{Shanghai Research Center for Quantum Sciences and CAS Center for Excellence in Quantum Information and Quantum Physics, University of Science and Technology of China, Shanghai, China}\affiliation{Hefei National Research Center for Physical Sciences at the Microscale and School of Physical Sciences, University of Science and Technology of China, Hefei, China}
\author{Qiang Zhang}\affiliation{Hefei National Laboratory, University of Science and Technology of China, Hefei, China}\affiliation{Shanghai Research Center for Quantum Sciences and CAS Center for Excellence in Quantum Information and Quantum Physics, University of Science and Technology of China, Shanghai, China}\affiliation{Hefei National Research Center for Physical Sciences at the Microscale and School of Physical Sciences, University of Science and Technology of China, Hefei, China}\affiliation{Jinan Institute of Quantum Technology and CAS Center for Excellence in Quantum Information and Quantum Physics, University of Science and Technology of China, Jinan, China}
\author{Yu-Ao Chen}\affiliation{Hefei National Laboratory, University of Science and Technology of China, Hefei, China}\affiliation{Shanghai Research Center for Quantum Sciences and CAS Center for Excellence in Quantum Information and Quantum Physics, University of Science and Technology of China, Shanghai, China}\affiliation{Hefei National Research Center for Physical Sciences at the Microscale and School of Physical Sciences, University of Science and Technology of China, Hefei, China}
\author{Hai-Feng Jiang}\affiliation{Hefei National Laboratory, University of Science and Technology of China, Hefei, China}\affiliation{Shanghai Research Center for Quantum Sciences and CAS Center for Excellence in Quantum Information and Quantum Physics, University of Science and Technology of China, Shanghai, China}\affiliation{Hefei National Research Center for Physical Sciences at the Microscale and School of Physical Sciences, University of Science and Technology of China, Hefei, China}
\author{Jian-Wei Pan}\affiliation{Hefei National Laboratory, University of Science and Technology of China, Hefei, China}\affiliation{Shanghai Research Center for Quantum Sciences and CAS Center for Excellence in Quantum Information and Quantum Physics, University of Science and Technology of China, Shanghai, China}\affiliation{Hefei National Research Center for Physical Sciences at the Microscale and School of Physical Sciences, University of Science and Technology of China, Hefei, China}
\date{\today}
\begin{abstract}
Fibre-linked optical clocks sense tidal changes of the gravitational potential, but separating overlapping tidal components and quantifying their precision remain challenging. Using strontium and ytterbium clocks 692 km apart, we measure over 353 days a response of 0.95 ± 0.11 to the combined solid-Earth and ocean-loading tide and 1.01 ± 0.13 for the lunar semidiurnal group, and separate the lunar and solar-frequency semidiurnal contributions. A signal-to-noise model based on the actual sampling accounts for the lunar precision and guides future clock-network observations.
\end{abstract}
\maketitle
Optical clocks connected by frequency-transfer links measure gravitational potential differences through the gravitational redshift. In the weak-field limit of general relativity, a potential difference \(\Delta U\) produces a fractional frequency shift \(\Delta\nu/\nu\simeq\Delta U/c^2\), where \(c\) is the speed of light.~\cite{r3} State-of-the-art optical clocks have reached evaluated fractional systematic uncertainties approaching \(10^{-19}\),~\cite{r10} and near Earth's surface fractional shifts of \(10^{-18}\), \(10^{-19}\) and \(10^{-20}\) correspond to equivalent heights of approximately 1 cm, 1 mm and 0.1 mm.~\cite{r5,r11} The redshift has been resolved within a millimetre-sized atomic sample,~\cite{r11} fibre networks and free-space frequency transfer have extended clock comparisons to distant sites,~\cite{r4,r12,r13,r14} and proposed quantum clock networks~\cite{r15} have reached an elementary demonstration with entangled clocks.~\cite{r16}

Clock comparisons complement levelling and satellite gravity observations.~\cite{r17,r18} However, Earth's gravitational potential also changes with volcanic activity,~\cite{r19} tsunamis~\cite{r20} and water storage,~\cite{r21} and these contributions overlap in a clock record. Detecting a weak signal therefore requires identifying the stronger variations and knowing how accurately each of them can be removed. Tides provide a predictable signal with which to develop this capability: the lunar and solar contributions must be distinguished from each other and from laboratory variations that follow the solar day, while observational gaps spread spectral power into neighbouring frequencies. Subtracting the tidal background thus requires separating the tidal components and their sampling aliases, and quantifying the precision with which each component is measured.

Here we report measurements of tidal changes in the potential difference between a strontium clock in coastal Shanghai~\cite{r6} and an ytterbium clock in inland Wuhan,~\cite{r7} located 692 km apart and connected by a 1,350-km fibre link.~\cite{r22} Over 545.9 hours of observations spanning 353 days, we measure the combined solid-Earth and ocean-loading tide and separate the principal lunar semidiurnal contribution (M\textsubscript{2}) from the contribution at the principal solar semidiurnal frequency (S\textsubscript{2}). The two contributions drift in relative phase and produce neighbouring spectral peaks, yet each remains identifiable after the other is removed. Furthermore, we develop a signal-to-noise model that combines signal strength, noise and actual sampling; it accounts for most of the differences in precision between components and provides a quantitative guide for future observations.

As shown in Fig. 1a, the Shanghai Sr clock and the Wuhan Yb clock lie on a 692-km coastal--inland baseline and are connected through Hefei by a 1,350-km phase-stabilized fibre link.~\cite{r22} For the spliced summer record (Ref.~\cite{r23}), the comparison instability is approximately \(5.4\times10^{-16}/\sqrt{\tau/\mathrm{s}}\) at averaging time \(\tau\) and is dominated by the clocks. The frequency ratio follows the changing potential difference, and we fit the observations as \(y(t)=c_j+A\,s(t)\), where \(s(t)\) is the predicted tidal change in fractional frequency, \(A\) is the response and the offset \(c_j\) of observing segment \(j\) absorbs the static part.

\begin{figure*}[t]
\includegraphics[width=\textwidth]{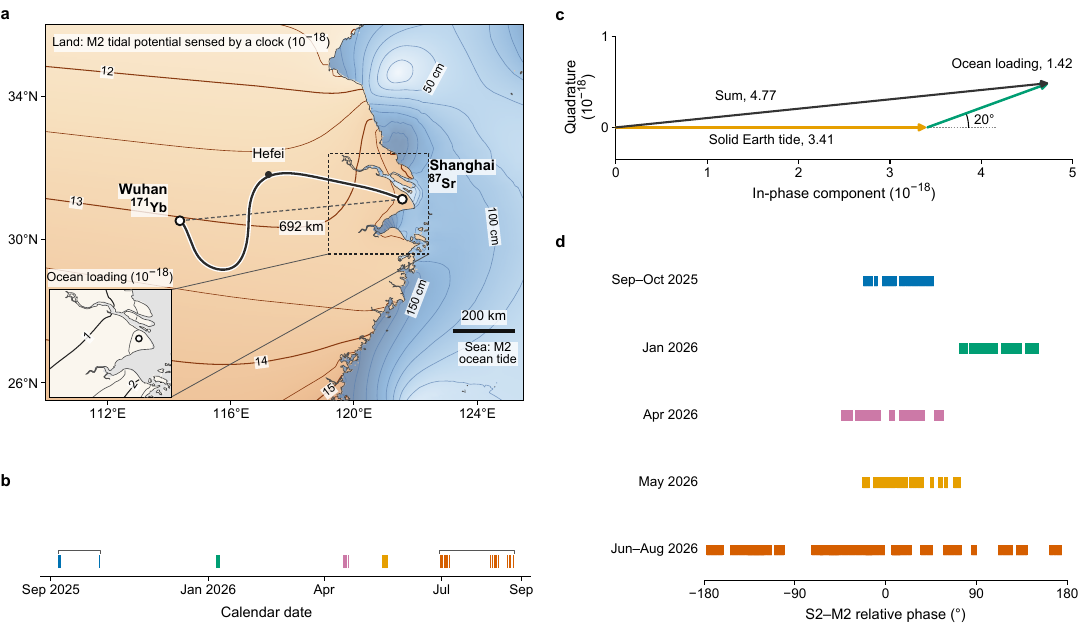}
\caption{\label{fig:1}\textbf{The predicted tide on a coastal--inland baseline.} (a) The Sr clock in Shanghai, the Yb clock in Wuhan and the fibre link through Hefei; the dashed line marks the 692-km baseline. Land contours show the predicted M\textsubscript{2} potential sensed by a clock, including the solid Earth tide and ocean loading, in units of 10\textsuperscript{−18} in fractional frequency (interval 0.5; integer contours labelled). Sea shading and contours show the sea-surface height amplitude of the FES2022 M\textsubscript{2} ocean tide, which drives the loading (interval 25 cm; 50, 100 and 150 cm labelled). The inset shows the ocean-loading contribution within the dashed box around Shanghai, with contours at 1 and 2 × 10\textsuperscript{−18}. (b) Retained observation intervals of the five campaigns, September 2025 to August 2026, in the filtered validation analysis; brackets span campaigns with separated runs. (c) Predicted differential M\textsubscript{2} phasors of the solid Earth tide, ocean loading and their sum, referenced to the solid Earth tide; the loading phasor is at 20°. (d) Relative astronomical phase of S\textsubscript{2} and M\textsubscript{2}, including the nodal correction, at the filtered 600-s samples. The summer campaign covers the full cycle; shorter campaigns cover limited ranges.}
\end{figure*}

On this baseline, ocean loading contributes appreciably. We compute the solid-Earth tide following the IERS conventions~\cite{r1} and the loading potential from FES2022,~\cite{r24,r25} including water attraction, induced Earth potential and clock displacement.~\cite{r2} As shown in Fig. 1c, the solid-Earth and loading phasors of the principal lunar semidiurnal constituent, M\textsubscript{2}, combine to a fractional-frequency amplitude of \(4.77\times10^{-18}\); in the physical prediction, each tidal constituent includes the combined solid-Earth and loading contribution at its frequency.

The observations comprise five campaigns and 54 observing segments over 353 days (Fig. 1b). The main analysis uses 545.9 h of retained samples to estimate template responses, spectra and signal-to-noise ratios, and a filtered analysis of the same observations provides the time-domain displays (Fig. 2) and validation tests. Both analyses fit segment offsets; their data selection, weighting and tidal-template realizations are described in Supplemental Material~\cite{SM} and Table S1.

\begin{figure*}[t]
\includegraphics[width=\textwidth]{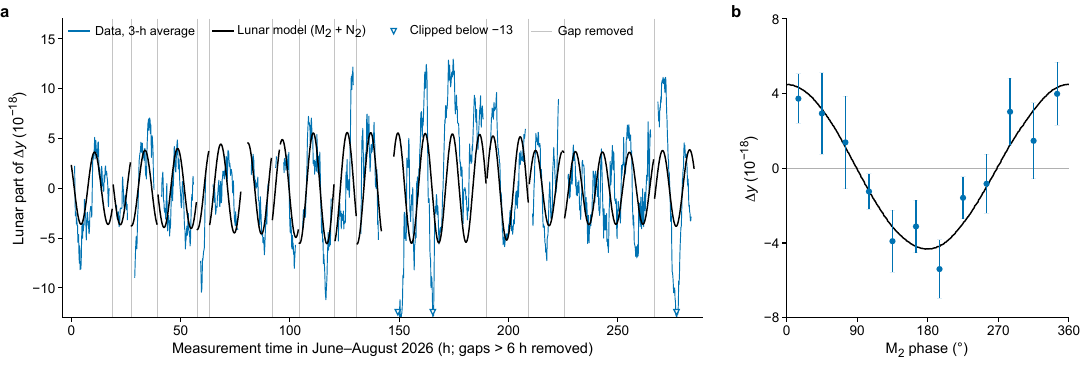}
\caption{\label{fig:2}\textbf{The lunar tide in the summer record.} Both panels use the filtered validation analysis (Supplemental Material,~\ref{sm:filtered-validation}~\cite{SM}). (a) Three-hour running averages of the June--August 2026 frequency ratio after subtracting fitted segment offsets and the fitted near-12-h and diurnal (non-lunar) tidal groups (blue), with the lunar model (black). The measurement-time axis compresses gaps longer than 6 h; grey lines mark their removal, and values below −13 × 10\textsuperscript{−18} are clipped (open triangles). White frequency noise alone corresponds to a scatter of approximately 5 × 10\textsuperscript{−18} for a 3-h average. (b) The same data folded at the model M\textsubscript{2} phase, relative to the predicted maximum, in 30° bins, with the lunar model (M\textsubscript{2} and N\textsubscript{2}) folded on identical sampling. Error bars are day-by-day jackknife uncertainties with the fitted non-lunar terms held fixed. No empirical daily terms are fitted.}
\end{figure*}

We obtain a full-template response of \(A=0.95\pm0.11\), where \(A=1\) denotes agreement with the predicted change in frequency ratio (Fig. 3a). To separate the semidiurnal contributions, we group M\textsubscript{2} with the larger lunar elliptic semidiurnal tide, N\textsubscript{2}, and the principal solar semidiurnal tide, S\textsubscript{2}, with the lunisolar semidiurnal tide, K\textsubscript{2}; within each group, relative amplitudes and phases follow the physical model. The lunar-group response is \(1.01\pm0.13\) (Fig. 3a), and when its phase is allowed to vary, the fitted timing is consistent with the prediction (Fig. 3b).

\begin{figure*}[t]
\includegraphics[width=\textwidth]{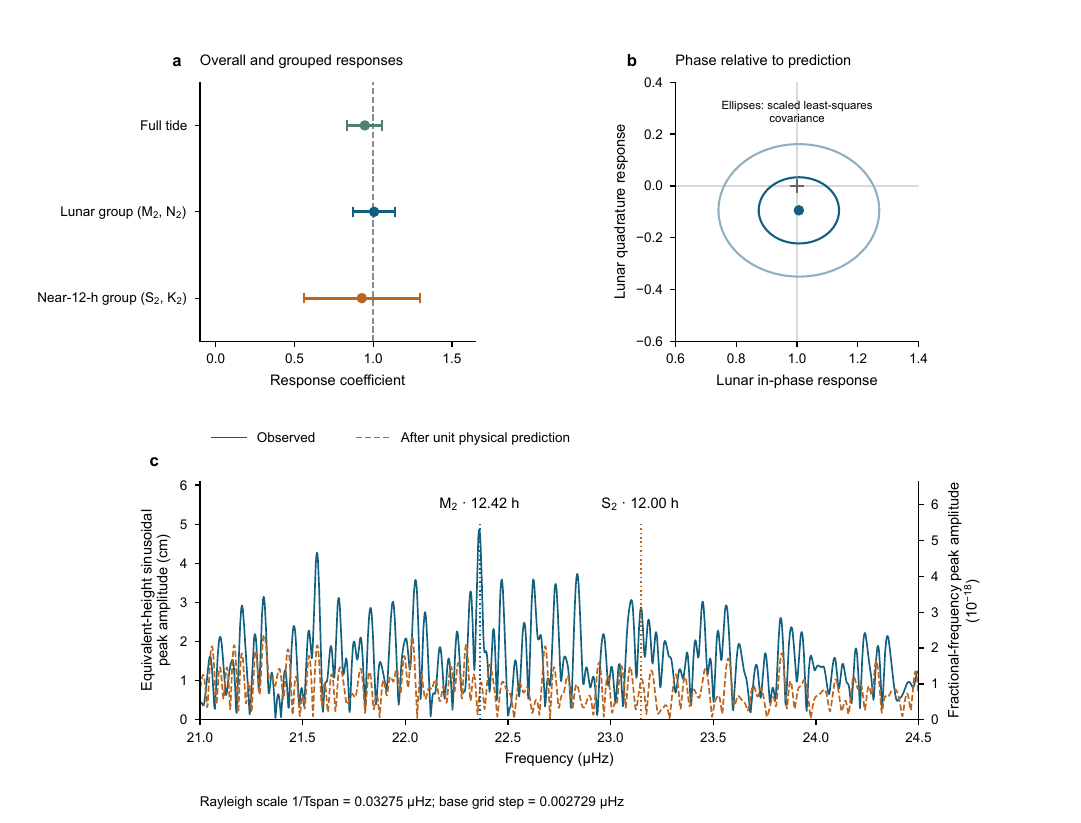}
\caption{\label{fig:3}\textbf{Amplitude, phase and frequency of the tidal response.} All panels use the main analysis: 3,299 non-overlapping 600-s means with 54 segment constants. (a) Complete-template (`full tide') response and jointly fitted lunar (M\textsubscript{2}, N\textsubscript{2}) and near-12-h (S\textsubscript{2}, K\textsubscript{2}) group responses, from separate fits with period-dependent noise weights. Error bars are one standard error, the larger of the scaled least-squares and UTC-day-deletion jackknife errors. (b) Lunar-group in-phase and quadrature coefficients; the cross marks \((1,0)\), and contours have radii one and two in the scaled least-squares covariance. The quoted marginal errors, the larger of the scaled least-squares and UTC-day-jackknife values, are 0.13 and 0.20; the scaled least-squares quadrature error is 0.128. (c) Equivalent-height sinusoidal peak-amplitude spectra over 21--24.5 μHz, before and after subtracting the fixed unit-amplitude physical prediction, with equal weights and segment constants projected from data and trial sinusoids. The left and right axes give equivalent height and fractional-frequency amplitude, related by \(A_{h,\mathrm{cm}}=100c^2A_y/g_0\) with standard gravity \(g_0=9.80665\ \mathrm{m/s^2}\); both are linear. Markers indicate the M\textsubscript{2} and S\textsubscript{2} frequencies; the nominal Rayleigh scale (0.03275 μHz) and grid step (0.002729 μHz) are annotated. Conditional spectra testing the separation of the M\textsubscript{2} and S\textsubscript{2}-frequency contributions are shown in Fig. S3c.}
\end{figure*}

The lunar and solar semidiurnal periods are approximately 12.42 h and 12 h, and their changing relative phase allows the observations to distinguish the two. The two-month summer campaign covers the full relative-phase cycle, whereas the shorter campaigns cover narrower intervals (Fig. 1d). As shown in Fig. 2, after removal of fitted offsets and non-lunar groups, the summer time series and its dependence on lunar phase agree well with the predicted variation. Campaign-resolved responses and records are given in Figs. S1 and S2, and the 274-h summer record was also used in the absolute frequency-ratio measurement of Hou et al.~~\cite{r23}

The equivalent-height amplitude spectrum shows lunar and solar-frequency peaks of \(4.9\pm0.7\) and \(2.9\pm1.1\) cm at 22.361 and 23.148 μHz, respectively (Fig. 3c; UTC-day-jackknife errors). Each principal frequency retains its peak when the other is controlled by a freely phased sinusoid, with N\textsubscript{2} and K\textsubscript{2} also controlled (Fig. S3). The conditional spectra thus separate the M\textsubscript{2} and S\textsubscript{2}-frequency contributions without imposing their astronomical phase relation. Note that the scan peaks are single-frequency amplitudes, whereas Table 1 uses joint M\textsubscript{2} and S\textsubscript{2} estimates that account for neighbouring constituents (Supplemental Material,~\ref{sm:spectra}~\cite{SM}; Table S2b). In contrast, the predicted-phase group responses in Fig. 3a test agreement with the physical waveforms: with the relative phases of S\textsubscript{2} and K\textsubscript{2} fixed within their group, the near-12-h group response is \(0.93\pm0.37\) (Supplemental Material,~\ref{sm:response-fits}~\cite{SM}).

Sampling aliasing in the finite, interrupted record introduces correlations between frequency components, which we test by reciprocal subtraction (Fig. S4); both contributions remain identifiable. The maximum canonical correlation between the projected sine--cosine pairs is 0.290 (Fig. S5), and the strongest sampling aliases respond to lunar-waveform subtraction as predicted (Fig. S6).

Furthermore, the lunar estimate is stable against specified daily disturbances. A \(10^{-18}\) fractional-frequency sinusoid at 24 h or 12 h changes the main lunar response by at most 0.025 when its phase is fixed across the record, and by at most 0.032 and 0.13, respectively, when the phases are independent between campaigns. The filtered analysis provides complementary checks of daily terms, sampling and data selection (Fig. 4).

\begin{figure*}[t]
\includegraphics[width=\textwidth]{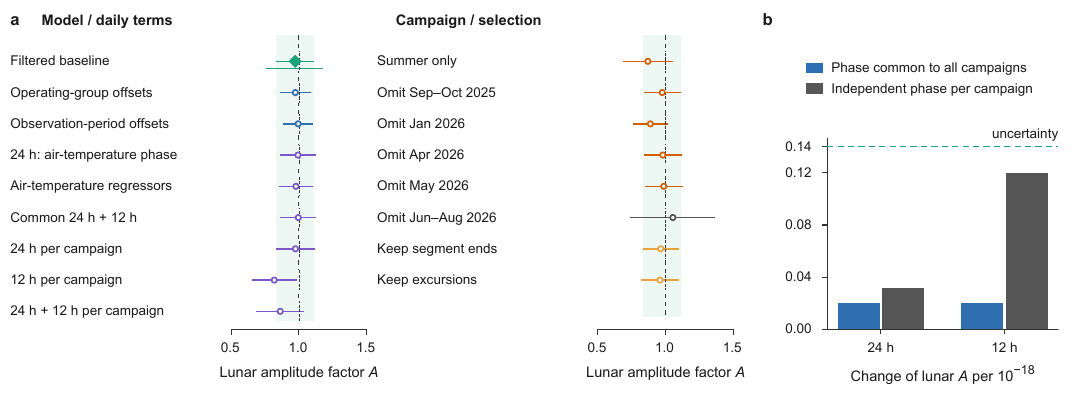}
\caption{\label{fig:4}\textbf{Robustness of the lunar amplitude.} Both panels use the filtered validation analysis; the corresponding main-analysis lunar response is 1.01 ± 0.13, from the same observing segments. (a) Lunar amplitude for the filtered baseline fit (0.97 ± 0.14; filled diamond and shaded uncertainty band), alternative offset definitions, daily terms represented by air-temperature measurements or fitted 24-h and 12-h sinusoids, the summer campaign alone, omission of each campaign, and retention of segment ends or transient excursions. Error bars and the shaded band indicate one standard deviation. The thin bar below the filtered baseline is the 95\% interval from resampling operating groups. The grey interval is obtained without the summer campaign. (b) Change in the lunar amplitude caused by a 24-h or 12-h sinusoid of fractional-frequency amplitude 10\textsuperscript{−18}, with one phase common to all campaigns or with the combination of independent campaign phases that produces the largest change. The dashed line marks the uncertainty of the filtered baseline estimate.}
\end{figure*}

The separate ocean-loading and diurnal responses are constrained less precisely, with signal-to-noise ratios near unity (Table 1). To quantify the precision of each component, we develop a signal-to-noise model that connects signal strength and noise to estimator precision through the actual sampling window, the fitted offsets and the coupling between components; within short segments, for example, a slowly varying tide is partly absorbed by the fitted offset. For a specified estimator with constant exponents, the amplitude signal-to-noise ratio scales as

\[\rho_i(T)\simeq\rho_i(T_0)
\frac{\eta_i(T)}{\eta_i(T_0)}
\left(\frac{T}{T_0}\right)^{a_i-b_i}.\tag{1}\]

Here \(T\) is the total retained exposure and \(T_0\) is the duration of one period of the target signal (its dominant period for a multi-constituent template); \(\rho_i=S/R\) is the ratio of the estimated amplitude \(S=|\hat A_i|\) to its standard error \(R=u(\hat A_i)\). The reference \(\rho_i(T_0)\) is the signal-to-noise ratio of one uninterrupted period in white noise, for which \(\eta_i(T_0)=1\) and \(\rho_i(T_0)=|\hat A_i|s_{i,\mathrm{rms}}\sqrt{T_0}/\alpha\), with the amplitude and template r.m.s. of the full record. The exponents \(a_i\) and \(b_i\) describe how the phase-aligned signal and the noise standard deviation accumulate with the number of periods, and \(\eta_i\) accounts for the sampling window, weights and parameter coupling of the actual estimator. For a coherent signal in white noise, \(a_i=1\) and \(b_i=1/2\), giving \(\rho_i\propto\sqrt{T}\) at fixed \(\eta_i\). We evaluate this white-noise case on the actual sampling with the measured amplitudes and a common clock-noise coefficient (Supplemental Material,~\ref{sm:signal-to-noise}~\cite{SM}), and compare the resulting precision with the empirical errors of the same estimators in Table 1.

\begin{table*}[t]
\caption{\label{tab:snr}Model-estimated and measured signal-to-noise ratios. \(S=|\hat A|\) is the measured amplitude or template-response magnitude and \(R=u(\hat A)\) its standard error. The model-estimated row assumes white frequency noise and uses the common noise coefficient and the actual sampling; the measured row uses UTC-day-jackknife covariance. All entries use the main analysis: M\textsubscript{2} and S\textsubscript{2} from their joint sine--cosine fit, the diurnal group from the three-group fit and ocean loading from the two-template solid-Earth/loading fit (Supplemental Material,~\ref{sm:signal-to-noise}~\cite{SM}).}
\begin{ruledtabular}
\begin{tabular}{lcccc}
Evaluation & Lunar M\textsubscript{2} & Solar-frequency S\textsubscript{2} & Diurnal group & Ocean loading \\
\hline
\(S/R\) (model estimated) & 7.17 & 2.78 & 1.50 & 1.31 \\
\(S/R\) (measured) & 7.02 & 1.96 & 1.03 & 1.00 \\
\end{tabular}
\end{ruledtabular}
\end{table*}

The model-estimated signal-to-noise ratio agrees well with the measured value for the lunar component, whereas the agreement is poorer for the weaker components (Table 1). For weak signals, residual sampling aliases (Figs. S5 and S6) and non-white laboratory variations following the solar day (Fig. S1d; Supplemental Material,~\ref{sm:daily-checks}~\cite{SM}) have a larger relative effect, particularly on the diurnal and solar-frequency components. In addition, short segments constrain diurnal cycles weakly after offset removal, and the smaller loading signal overlaps the solid-Earth waveform. The signal-to-noise model thus accounts for most of the differences in precision between components, evaluated at the measured amplitudes.

The model also identifies distinct ways to improve future observations. Better clock stability lowers the noise, whereas longer uninterrupted records reduce the loss of slowly varying signals to fitted offsets. In an idealized 96-h record with white frequency noise, extending the offset interval from 12 h to 24 h reduces the phase-averaged uncertainty for an isolated K\textsubscript{1} diurnal sinusoid by 43\%, with no further gain at longer intervals (Supplemental Material,~\ref{sm:sampling-diagnostics}~\cite{SM}). Covering complete diurnal cycles therefore improves precision even at fixed exposure. A longer calendar span separates neighbouring frequencies, while gaps create sampling aliases (Fig. S6). Together with increased exposure, additional sites and local environmental records, these improvements would help identify weaker potential changes after subtraction of the tidal background.

The measurement and separation of the lunar and solar-frequency semidiurnal contributions extend demonstrated sub-centimetre-equivalent clock performance~\cite{r5} to time-varying geophysical signals and, together with the signal-to-noise model, connect component separation to the design of future clock observations. Characterizing the tidal background has substantial applications to the redefinition of the second~\cite{r8} and the unification of height systems at the centimetre level.~\cite{r9}

\textit{Note added.}~During the preparation of this manuscript, a related observation of tidal variations of the gravity potential in a European optical-clock network was reported;~\cite{r26} the present work is complementary in separating the lunar and solar-frequency semidiurnal contributions on a coastal--inland baseline.
\begin{acknowledgments}
This work was supported by the National Natural Science Foundation of China (grant nos. T2125010, 92476001 and 12374467); the Quantum Science and Technology-National Science and Technology Major Project (grant nos. 2021ZD0300105, 2021ZD0300106, 2021ZD0300902, 2021ZD0300903 and 2021ZD0300904); the Strategic Priority Research Program of the Chinese Academy of Sciences (grant nos. XDA0520101, XDA0520102, XDA0520103 and XDB35000000); the Hefei National Facility for Quantum Precision Measurement (https://cstr.cn/31140.02.HQPM); the Independent Deployment Project of HFNL (grant nos. ZB2025010400 and ZB2025010100); the Anhui Initiative in Quantum Information Technologies (grant nos. AHY010100 and AHY040200); the Shanghai Municipal Science and Technology Major Project (grant no. 2019SHZDZX01); the Scientific Research Innovation Capability Support Project for Young Faculty (grant no. ZYGXQNJSKYCXNLZCXM-I26); the National Science and Technology Major Project (grant no. 2025ZD0300702); and the New Cornerstone Science Foundation. Large language models were used to assist with code writing, manuscript drafting and editing. All scientific content, analyses and conclusions are those of the authors, who independently verified all numerical results and take full responsibility for the manuscript.
\end{acknowledgments}

\clearpage
\onecolumngrid
\setcounter{figure}{0}\setcounter{table}{0}\setcounter{equation}{0}\setcounter{section}{0}
\renewcommand{\thefigure}{S\arabic{figure}}
\renewcommand{\thetable}{S\arabic{table}}
\renewcommand{\theequation}{S\arabic{equation}}
\begin{center}{\large\bfseries Supplemental Material}\end{center}
\vspace{1em}
\setcounter{secnumdepth}{1}
\setcounter{section}{0}
\renewcommand{\thesection}{SM\arabic{section}}
\makeatletter
\renewcommand{\p@section}{}
\makeatother
\section{Measurements and record construction}\label{sm:measurements}
The Shanghai Sr clock (31.130° N, 121.570° E) and Wuhan Yb clock (30.529° N, 114.357° E) have evaluated systematic uncertainties of \(9.2\times10^{-19}\) and \(1.3\times10^{-18}\), respectively.~[1,2] A 1,550-nm carrier referenced to Sr through a frequency comb is transmitted to Wuhan, where its beat against the Yb-referenced comb is counted at 1-s intervals.~[3] In 2026 the beat is 33.6 MHz, with 1 Hz corresponding to \(5.17\times10^{-15}\) in Yb/Sr; in 2025 the counter recorded the beat after division by two, which has been taken into account. Fitted offsets absorb the static geopotential difference. Monitoring the out-of-loop round-trip signal and optical phase-locked loops checks the phase connection needed for coherent accumulation.~[3]

Five campaigns span September 2025--August 2026. The 54 experiment-defined segments include 38 before June and 16 summer segments from Ref.~[3]. Retained one-second samples in the main analysis total 545.9 h. Campaigns define seasonal subsets and campaign-specific daily terms; eight observing periods define the noise weights and alternative offsets, 29 operating groups define group-bootstrap resampling and alternative offsets, and segments carry the baseline fitted offsets. Time stamps are local time (UTC+8).

The main analysis (auxiliary reduction) supplies responses, spectra and signal-to-noise ratios; filtered validation (primary reduction) supplies time-domain displays and robustness checks. They share observations and are not independent measurements (Table S1). The main analysis rejects records more than 10 Hz from each segment median and forms 3,299 non-overlapping 600-s means on a local wall-clock grid, retaining bins with at least 180 nominal one-second members. Timestamps are nominal grid centres.

Filtered sample selection and transient screening are detailed in Supplemental Material,~\ref{sm:filtered-validation}. Screening excludes 0.4\% of samples, or 2.2 h. The main analysis applies neither this transient screening nor the triangular filter below.

\section{Physical predictions}\label{sm:physical-predictions}
Both analyses use the IERS 2010 solid-Earth tide with constituent-dependent Love numbers,~[4] evaluated at bin centres for the main analysis. At each clock,

\[\delta W=\sum_m(1+k_{2m})V_{2m}+(1+k_3)V_3-g u_r,\tag{S1}\]

where \(V_{2m}\) are the lunar and solar degree-2 tide-generating potentials, \(V_3\) is the lunar degree-3 potential and \(u_r\) is the IERS radial displacement. The ocean-loading potential is

\[\delta W_{\rm ocean}(t)=\sum_{\rm ocean}\rho_w\zeta(\mathbf r',t)G_W(\psi)\,\mathrm dA,
\qquad
G_W(\psi)=\frac{ga}{M_E}\sum_n(1+k'_n-h'_n)P_n(\cos\psi),\tag{S2}\]

using FES2022 elevations \(\zeta\) on a \(1/30^\circ\) grid,~[5,6] \(\rho_w=1,025\ \mathrm{kg\,m^{-3}}\), Earth's radius \(a\) and mass \(M_E\), angular distance \(\psi\), and PREM load Love numbers.~[7--9] The main loading template uses 18 constituents. The filtered model synthesizes eight constituents with astronomical arguments and nodal corrections after removing the ocean mean; remaining terms are held at their prediction. Extending the ocean mask near Shanghai by 0.3° or 0.6° changes the filtered lunar response by −0.05 or −0.08, about half its statistical uncertainty of 0.14. Implementation and loading validation are given in Supplemental Material,~\ref{sm:loading-validation} and Fig. S7.

\section{Response fits and uncertainties}\label{sm:response-fits}
The lunar (M\textsubscript{2}, N\textsubscript{2}), near-12-h (S\textsubscript{2}, K\textsubscript{2}) and diurnal (K\textsubscript{1}, O\textsubscript{1}, P\textsubscript{1}, Q\textsubscript{1}) templates are \(s_L\), \(s_S\) and \(s_D\); other terms form a fixed remainder \(r_0\). The complete-template and grouped models are

\[\begin{aligned}
y_i&=c_{s(i)}+A_T(s_{L,i}+s_{S,i}+s_{D,i}+r_{0,i})+\epsilon_i,\\
y_i-r_{0,i}&=c_{s(i)}+A_Ls_{L,i}+A_Ss_{S,i}+A_Ds_{D,i}+\epsilon_i.
\end{aligned}\tag{S3}\]

Here \(y_i\) is the fractional-frequency mean of bin \(i\), \(c_{s(i)}\) is the offset of its segment \(s(i)\) and \(\epsilon_i\) is the residual.

The main phase-relaxed fit adds a quarter-cycle-shifted lunar template, preserving the M\textsubscript{2}/N\textsubscript{2} relation. Main-response weights are \(W_{ii}=1/\sigma_i^2\), with \(\sigma_i=\alpha_{p(i)}/\sqrt{\Delta t_i}\) and retained bin exposure \(\Delta t_i=n_i\delta t\), where \(\delta t=1\) s. Period coefficients \(\alpha_p\) come from overlapping Allan deviations at 256--4,096 s:~[10] \(9.70\times10^{-16}\) and \(5.70\times10^{-16}\ \sqrt{\mathrm{s}}\) for September and October 2025, and \((4.30\)--\(5.76)\times10^{-16}\ \sqrt{\mathrm{s}}\) for six 2026 periods. Errors are the larger of residual-scaled least-squares~[11] and UTC-day-deletion jackknife errors, refitting all coefficients at each deletion.

Filtered observations and models receive identical 1,200-s triangular filtering at 600-s steps. Equal-weight fits use segment offsets and either a complete template, semidiurnal and diurnal bands, or the three groups above; the baseline grouped fit has no empirical daily terms. Quoted fit uncertainties are statistical standard deviations; resampling and display construction are detailed in Supplemental Material,~\ref{sm:filtered-validation}.

\section{Spectra and constituent separation}\label{sm:spectra}
For Fig. 3c, equal-weight fits of the main means at each frequency use

\[y_i=c_{s(i)}+\beta_{c,f}\cos[2\pi f(t_i-t_0)]+\beta_{s,f}\sin[2\pi f(t_i-t_0)]+\epsilon_i.\tag{S4}\]

Here \(t_0\) is 6 September 2025, 16:05 UTC. The peak amplitude is \(A_y(f)=\sqrt{\beta_{c,f}^2+\beta_{s,f}^2}\) and equivalent height \(A_{h,\mathrm{cm}}=100c^2A_y/g_0\), with standard gravity \(g_0=9.80665\ \mathrm{m\,s^{-2}}\); using normal gravity at the two site latitudes would increase the heights by approximately 0.13\%. Peak-amplitude errors are UTC-day-deletion jackknife errors, which exceed the scaled least-squares errors. The residual curve first subtracts the fixed unit-amplitude physical prediction. With \(T_{\rm span}=353.3889\) d, the scan spacing is \(1/(12T_{\rm span})\), adding astronomical frequencies; peaks are local maxima within \(1/T_{\rm span}\) of prescribed frequencies.

For Figs. S3--S6, nuisance terms \(N\) comprise segment constants and free N\textsubscript{2}/K\textsubscript{2} pairs where stated. An orthonormal basis \(Q\) defines \(R_N=I-QQ^{\mathsf T}\), giving \(\hat{\boldsymbol\beta}_f=(R_NX_f)^+R_Ny\) for trial cosine--sine pairs \(X_f\). Controlled frequencies lie in the nuisance space and have undefined amplitudes. Phases refer to \(t_0\). M\textsubscript{2} and S\textsubscript{2} are fitted jointly without relative amplitude or phase constraints (Table S2b). Conditional scans control the other principal pair within ±0.16 μHz of the target. Compensation subtracts a scaled donor waveform before refitting the target. Canonical correlations are singular values of \(Q_M^{\mathsf T}Q_S\), where \(Q_M\) and \(Q_S\) are orthonormal bases of the residualized pairs (Fig. S5).

\section{Signal-to-noise calculation}\label{sm:signal-to-noise}
We define \(\rho_i=|\hat A_i|/u(\hat A_i)\). For design \(X\) and weights \(W\), \(H=(X^{\mathsf T}WX)^{-1}X^{\mathsf T}W\) maps observations to coefficients. White frequency noise with \(\alpha=5.4\times10^{-16}\sqrt{\mathrm{s}}\) gives bin variance \(\alpha^2/\Delta t_k\) and

\[u_{i,\mathrm{white}}^2=\alpha^2\sum_k\frac{H_{ik}^2}{\Delta t_k},
\qquad
\rho_{i,\mathrm{white}}=\frac{|\hat A_i|}{u_{i,\mathrm{white}}}.\tag{S5}\]

For retained exposure \(T=\sum_k\Delta t_k\), template \(s_i\) per unit response, \(s_{i,\mathrm{rms}}^2=\sum_k\Delta t_k s_i(t_k)^2/T\) and \(\eta_i=\alpha/(s_{i,\mathrm{rms}}\sqrt T\,u_{i,\mathrm{white}})\),

\[\rho_{i,\mathrm{white}}=\frac{|\hat A_i|s_{i,\mathrm{rms}}}{\alpha}\eta_i\sqrt T.\tag{S6}\]

Sampling, offsets, coupling and weights enter \(\eta_i\); Eq. (S6) is Eq. (1) of the main text with \(a_i=1\), \(b_i=1/2\) and the one-period reference, evaluated at the measured amplitude (Table S2c). Table 1 uses equal-weight joint M\textsubscript{2}/S\textsubscript{2}/N\textsubscript{2}/K\textsubscript{2} pairs and 54 offsets for M\textsubscript{2}/S\textsubscript{2}, with \(\hat A=\sqrt{\beta_c^2+\beta_s^2}\) and \(u^2(\hat A)=JC_iJ^{\mathsf T}\), \(J=(\beta_c,\beta_s)/\hat A\), where \(C_i\) is the \(2\times2\) block of \(C_\beta\) for \((\beta_c,\beta_s)\). The white-noise covariance is \(C_\beta=H\Sigma H^{\mathsf T}\), \(\Sigma_{kk}=\alpha^2/\Delta t_k\); measured covariance comes from 43 UTC-day deletion fits. The diurnal column uses the three-group fit; loading uses equal-weight complete solid-Earth/loading templates with two coefficients, 54 offsets and UTC-day-deletion errors.

\section{Daily terms and environmental checks}\label{sm:daily-checks}
Filtered sensitivity fits add 24-h and 12-h sine--cosine pairs in local time, with common or campaign-specific coefficients. Alternatives use a free-amplitude 24-h term phased to each campaign's hourly ERA5 2-m air-temperature harmonic,~[12] or two site-temperature anomalies relative to a 24-h running mean.

For lunar estimator row \(\mathbf h_L\) of \(H\), a disturbance \(B\cos(\omega t-\varphi)\), with \(\omega\) the angular frequency of the 24-h or 12-h disturbance, has maximum common-phase effect

\[B\sqrt{(\mathbf h_L\cdot\cos\omega t)^2+(\mathbf h_L\cdot\sin\omega t)^2}.\tag{S7}\]

Independent campaign phases give the sum of campaign norms. Results use the main three-group design and period-noise weights; Fig. 4 uses the equal-weight filtered design.

The Shanghai round-trip link beat near 10 MHz, recorded every second, is normalized by the optical carrier and filtered like the ratio. Over 474 h, adding it as a regressor or excluding 300-s blocks deviating by more than \(5\times10^{-16}\) changes lunar and full-template responses by less than 0.005.

The mean of 20 Sr-chamber PT100 sensors sampled every 20 s covers 28 June--7 July 2026, overlapping 98 h of filtered summer data. Blackbody sensitivity is \(6.26\times10^{-17}\ \mathrm{K^{-1}}\) at 294.3 K. Both clocks correct blackbody shifts in real time, so regression on the chamber-temperature shift tests residual errors with expected coefficient zero; it gives a coefficient of \(0.92\pm0.74\). Fig. S1d shows the folded residuals and the chamber-derived blackbody-radiation shift.

\section{Physical model and loading validation}\label{sm:loading-validation}
We use \(k_{20}=0.29525\), \(k_{22}=0.29801\) and \(k_3=0.093\). The IERS 2010 resonance formula gives constituent-dependent \(k_{21}\) values of 0.2588 for K\textsubscript{1}, 0.2873 for P\textsubscript{1} and 0.2975 for O\textsubscript{1}. Radial displacement follows the IERS algorithm, including the frequency and latitude dependence of \(h\).~[4]

PREM load Love numbers through degree \(n=696\)~[8] come from gravity-toolkit.~[9] Above this degree, \(h'_n\) joins smoothly to its Boussinesq limit of −6.22; a Kummer transformation sums the series through \(n=40,000\). Full grid resolution is used within 3° of each station and \(8\times8\) cell blocks farther away, changing differential M\textsubscript{2} and K\textsubscript{1} loading by at most 0.05\%.

The full loading potential exceeds the vertical-displacement contribution \(-g\delta h\) by factors of approximately 1.4 for semidiurnal and 1.8 for diurnal constituents. Calculated radial displacement agrees with the FES2022 load-tide grid within 2.2\% in amplitude and 1.2° in phase for seven constituents (Fig. S7). The S\textsubscript{2} grid comes from a different ocean-model release and differs by about 18\% in amplitude; substituting its inferred loading changes the full-template response by 0.001 and leaves the lunar response unchanged.

Shanghai is 22 km from the nearest ocean cell. The coastal-mask check fills land cells within 0.3° or 0.6° of the coastline in a \(4^\circ\times5^\circ\) box centred on Shanghai with nearest-ocean values; the resulting response changes are given in Supplemental Material,~\ref{sm:physical-predictions}.

\section{Filtered validation, screening and uncertainty calibration}\label{sm:filtered-validation}
For filtered validation, time is reconstructed from the first timestamp and sample index. After the same 10-Hz rejection, the longest uninterrupted run is retained; leading and trailing samples differing from the median by more than 1\% of the peak-to-peak range are trimmed. Means over 300 s are compared with a running median over ±30 min; those beyond six robust standard deviations and their neighbours are removed.

Adding the tidal model with factors of −1, 1 or 3 leaves the removed samples unchanged. Relative to the filtered baseline, block lengths of 60 or 300 s and thresholds of four to eight standard deviations change the lunar response by −0.04 to +0.05, and retaining excursions changes it by −0.01. Before transient removal, the procedure reproduces the summer sample set of Ref.~[3].

For the full, semidiurnal and lunar responses, fitted residuals are shifted circularly within each continuous filtered run, added to the model with \(A=1\) and refitted. The standard deviation of 200 realizations calibrates the error. Diurnal and near-12-h responses, daily-term fits and daily amplitudes use the larger of scaled least-squares and day-by-day jackknife errors. Group-bootstrap intervals resample 29 operating groups with replacement in 1,000 refits; segment bootstraps resample 54 segments.

Lunar-group residual resampling gives the quoted standard deviation of 0.14, compared with 0.13 for the larger of scaled least-squares and jackknife errors. The five lunar campaign estimates are consistent (\(\chi^2=4.1\), four degrees of freedom, \(p=0.40\); Table S2a), as are the semidiurnal estimates (\(\chi^2=3.2\), \(p=0.52\); Fig. S1a).

For Fig. 2b, observations minus fitted offsets and non-lunar groups, and the M\textsubscript{2}+N\textsubscript{2} model, are averaged in 30° M\textsubscript{2}-phase bins on identical sampling. Day-by-day jackknife errors hold non-lunar fits fixed. Time-series displays use 3-h averages and mark compression of gaps longer than 6 h.

\section{Separation and sampling diagnostics}\label{sm:sampling-diagnostics}
For donor \(D\) and target \(T\), the joint fit supplies a fixed donor coefficient vector \(\hat\beta_D\). For 61 values of \(\lambda\) from 0 to 1.5, we form \(y_\lambda=y-\lambda X_D\hat\beta_D\) and refit only the target pair with nuisance terms \(N\). With \(P_T=R_NX_T\) and \(P_D=R_NX_D\),

\[\hat\beta_T(\lambda)=\hat\beta_{T,\mathrm{single}}-\lambda P_T^+P_D\hat\beta_D,\]

where \(P_T^+\) is the Moore--Penrose pseudoinverse. Amplitude is the coefficient-vector norm and phase is \(\operatorname{atan2}(\beta_s,\beta_c)\), corresponding to \(A\cos[2\pi f(t-t_0)-\phi]\). Equivalent-height conversion is defined in Supplemental Material,~\ref{sm:spectra}.

Template lag correlations evaluate the donor at \(t-\ell\) and target at \(t\), residualize on the observation support and normalize their inner product. Lags span ±24 h in 5-min steps. Joint-coefficient correlations use the scaled least-squares and UTC-day-deletion covariances (Fig. S5).

Alias diagnostics scan 20--25 μHz at spacing \(1/(12T_{\rm span})\). For each donor, the three strongest local maxima of its pure-injection least-squares power are selected at least \(2/T_{\rm span}\) from its own frequency. The fitted waveforms and their subtraction use the same sampling and nuisance projection (Figs. S3 and S6).

In the main two-template solid-Earth/loading fit, the projected template correlation is 0.720 and the coefficient correlation under day deletion is −0.887. These quantify template overlap and fitted-coefficient coupling, respectively.

To isolate the precision cost of segment offsets, we evaluate an idealized 96-h record of 576 fully exposed, non-overlapping 600-s means with independent white frequency noise and the noise coefficient of Eq. (S5). For each of M\textsubscript{2} and K\textsubscript{1} separately, \(X\) contains a free sine--cosine pair averaged over each bin; \(Z\) contains offsets at intervals of 6, 12, 24, 48 or 96 h. With \(R_Z=I-Z(Z^{\mathsf T}Z)^{-1}Z^{\mathsf T}\), the pair covariance is

\[C=\frac{\alpha^2}{600\,\mathrm{s}}(X^{\mathsf T}R_ZX)^{-1}.\]

We summarize it by \(u_\phi=\sqrt{\operatorname{tr}(C)/2}\), the square root of the variance averaged over directions in the sine--cosine plane. Extending the offset interval from 12 to 24 h reduces \(u_\phi\) for K\textsubscript{1} from \(2.292\times10^{-18}\) to \(1.299\times10^{-18}\), with no further change at 48 or 96 h; for M\textsubscript{2} the change from 12 to 48 h is below 0.01\%. The comparison holds exposure, sampling and noise fixed and treats each constituent separately under white noise.

\clearpage
\begin{figure*}[p]
\includegraphics[width=\textwidth]{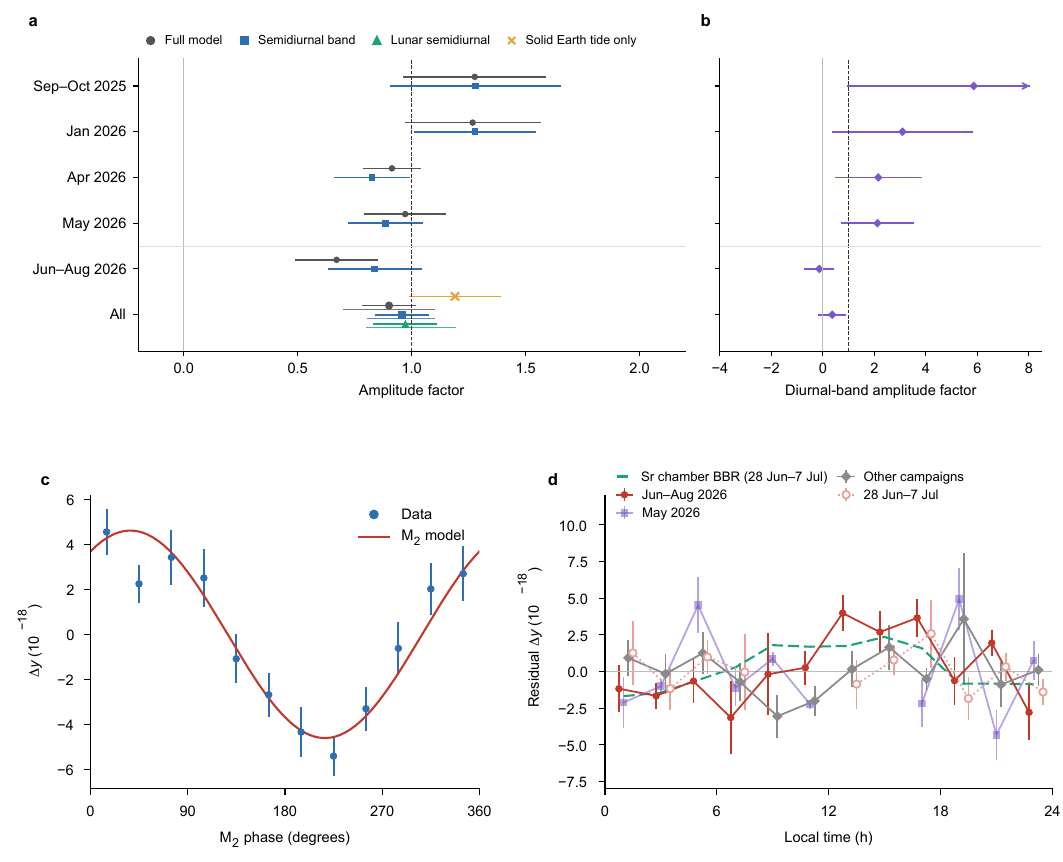}
\caption{\label{figureS1}\textbf{Campaign-resolved responses and phase folds.} All panels use the filtered validation analysis. (a) Full-model and semidiurnal responses for each campaign and for all data, fitted with segment offsets; the combined rows also show the lunar group and the solid Earth tide alone. (b) Diurnal response fitted jointly with the semidiurnal band; the combined estimate is 0.36 ± 0.50, and the arrow marks an error bar extending beyond the plotted range. Error bars are one standard deviation, calibrated by residual resampling for the full model, semidiurnal band and lunar group, and otherwise the larger of the scaled least-squares and day-by-day jackknife errors. Thin lines in the combined rows are 95\% intervals from resampling operating groups, or segments for the lunar group. (c) All observations folded at the M\textsubscript{2} phase in 30° bins after subtraction of fitted offsets and all constituents except M\textsubscript{2}; the curve is the M\textsubscript{2} model. (d) Tidally corrected residuals folded by local time in 2-h bins for summer, May (light symbols), the other three campaigns and the summer data overlapping the 28 June--7 July temperature record (open circles); bins covering fewer than three days are omitted. Folded points have day-by-day jackknife errors with removed terms held fixed. The dashed curve is the mean daily blackbody-radiation shift calculated from the Sr chamber temperature over the same interval, after removal of a 24-h running mean and the 30 June transient; the real-time correction removes this shift from the recorded ratio.}
\end{figure*}

\begin{figure*}[p]
\includegraphics[width=\textwidth]{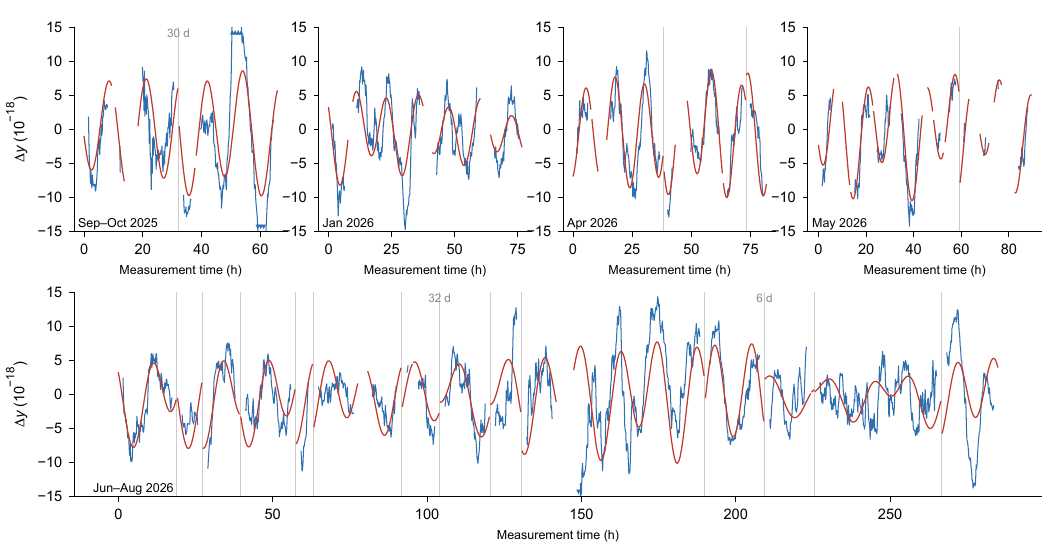}
\caption{\label{figureS2}\textbf{Tidal records across campaigns.} Three-hour averages of the filtered frequency ratio after subtraction of fitted segment offsets (blue), with the tidal model (red). Measurement-time axes compress gaps longer than 6 h, marked by vertical lines; gaps longer than 5 days are labelled. The vertical range is ±15 × 10\textsuperscript{−18}; triangles mark values outside this range.}
\end{figure*}

\begin{figure*}[p]
\includegraphics[width=\textwidth]{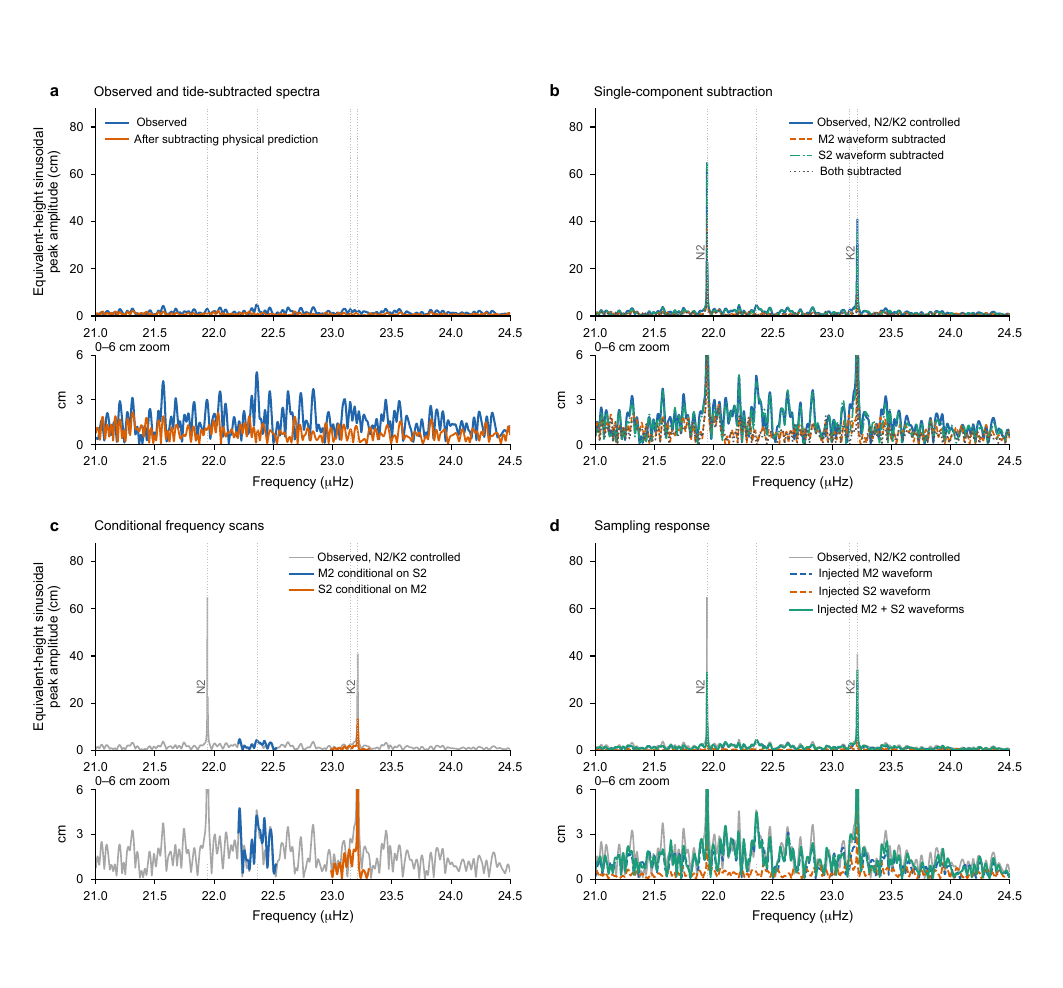}
\caption{\label{figureS3}\textbf{Amplitude spectra and sampling-response checks.} All panels use the main analysis with the 3,299 recorded bin-centre times and their gaps, equal weights and 54 segment constants. Ordinates give equivalent-height peak amplitude in cm on linear scales, with 0--6 cm zooms below each panel. (a) Observed amplitude spectrum and amplitude spectrum after subtraction of the fixed unit-amplitude physical prediction, as in Fig. 3c. (b) Amplitude spectra before and after subtraction of the fitted M\textsubscript{2}, S\textsubscript{2} or both waveforms, obtained from one simultaneous freely phased fit with N\textsubscript{2} and K\textsubscript{2} controlled. (c) Local scans around M\textsubscript{2} conditioned on S\textsubscript{2}, and around S\textsubscript{2} conditioned on M\textsubscript{2}, with N\textsubscript{2} and K\textsubscript{2} controlled; grey is the baseline from b. (d) M\textsubscript{2}-only, S\textsubscript{2}-only and combined fitted-waveform injections through the same sampling and projection, compared with the observed spectrum; injected amplitudes need not add. Dotted lines mark N\textsubscript{2}, M\textsubscript{2}, S\textsubscript{2} and K\textsubscript{2} in ascending frequency. Amplitudes at the controlled N\textsubscript{2} and K\textsubscript{2} frequencies are undefined, and responses close to them are weakly constrained.}
\end{figure*}

\begin{figure*}[p]
\includegraphics[width=\textwidth]{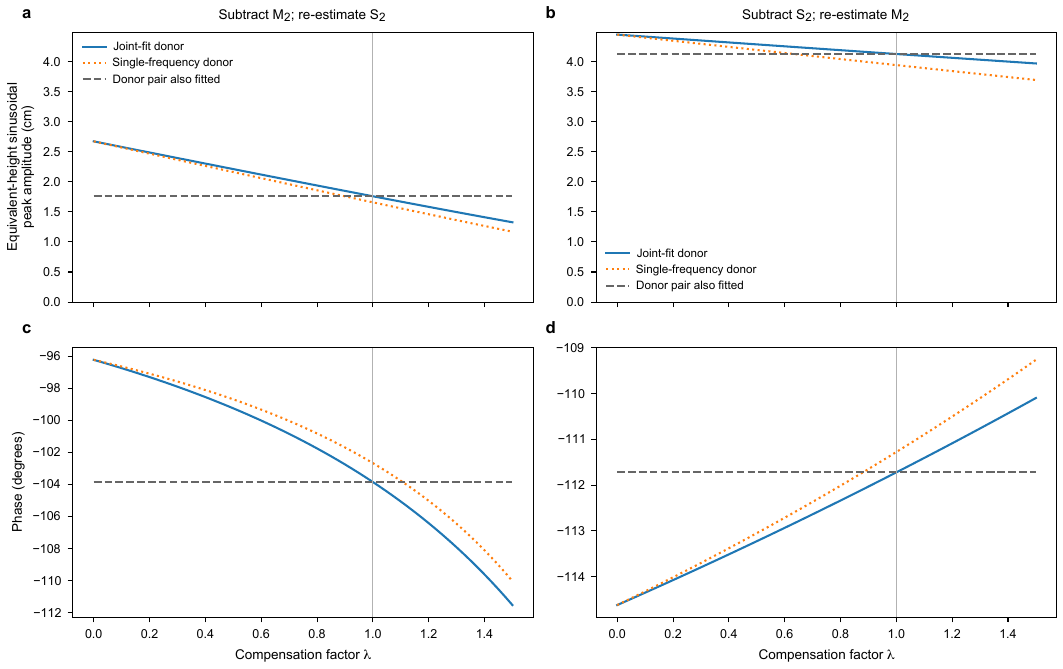}
\caption{\label{figureS4}\textbf{Amplitude and phase sensitivity to compensation.} (a,c) Subtracting a scaled M\textsubscript{2} waveform and re-estimating S\textsubscript{2}; (b,d) the reverse. (a,b) Equivalent-height peak amplitude, converted as in Fig. 3c; (c,d) phase relative to the common time origin. Solid curves use donor waveforms from the joint fit; dotted curves use single-frequency donor fits. All target-only fits retain segment offsets and free N\textsubscript{2}/K\textsubscript{2} pairs. Dashed grey curves (donor pair also fitted) are invariant by construction. λ = 1 denotes subtraction of the estimated waveform. Amplitude axes are linear and share a scale across directions.}
\end{figure*}

\begin{figure*}[p]
\includegraphics[width=\textwidth]{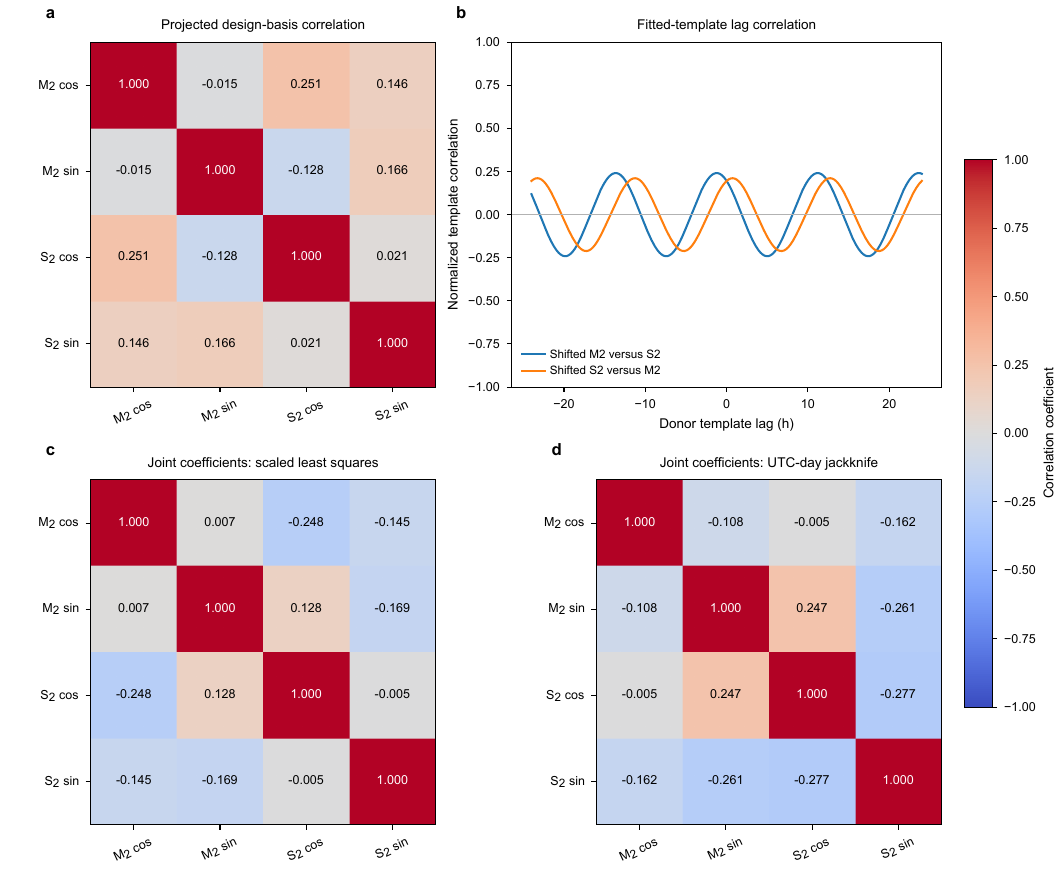}
\caption{\label{figureS5}\textbf{Harmonic geometry and joint-coefficient correlations in the main analysis.} (a) Correlations of the M\textsubscript{2} and S\textsubscript{2} cosine and sine design columns after projecting out 54 segment constants and free N\textsubscript{2}/K\textsubscript{2} pairs. (b) Normalized lag correlations of the fitted templates, evaluated on the original timestamps without interpolating missing observations. c,d, Correlations between the four jointly estimated coefficients (M\textsubscript{2} cosine, M\textsubscript{2} sine, S\textsubscript{2} cosine, S\textsubscript{2} sine) from the scaled least-squares covariance (c) and the UTC-day delete-one jackknife covariance (d). The three matrices share a −1 to +1 colour scale. The maximum canonical correlation of the projected M\textsubscript{2} and S\textsubscript{2} pairs is 0.290.}
\end{figure*}

\begin{figure*}[p]
\includegraphics[width=\textwidth]{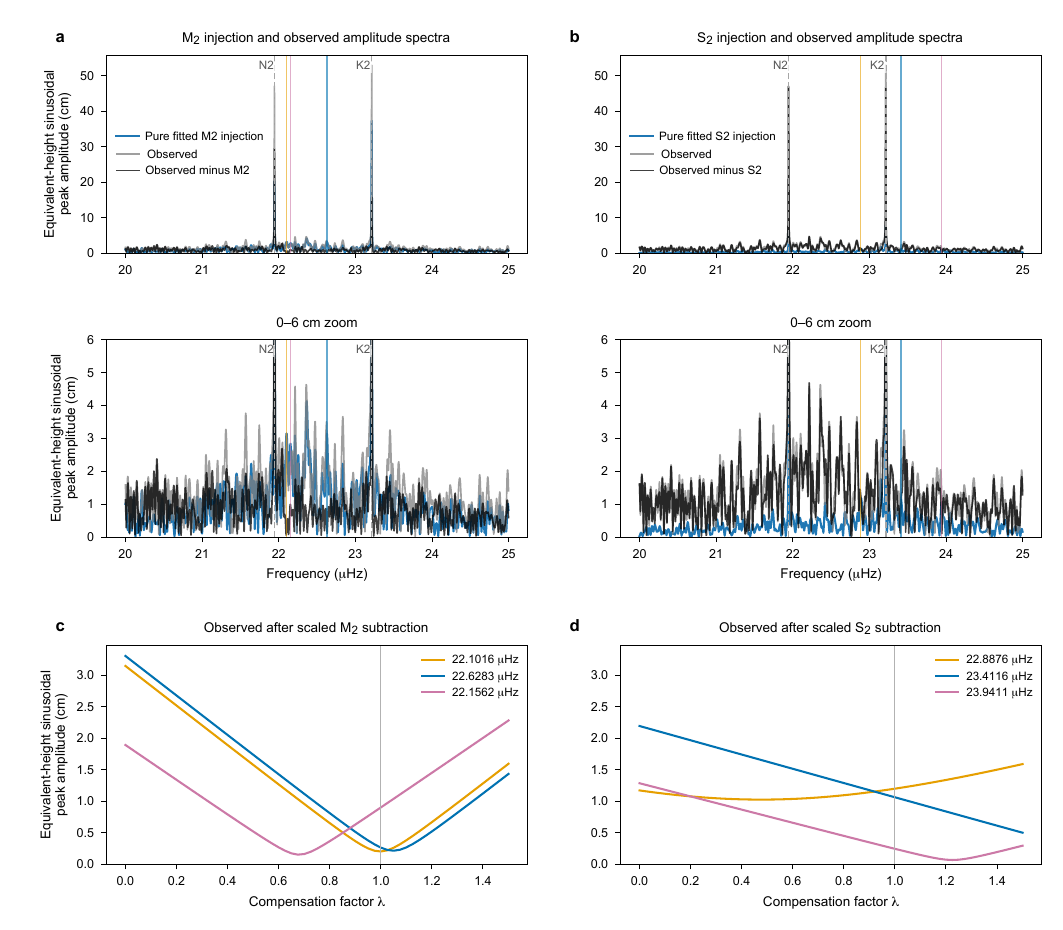}
\caption{\label{figureS6}\textbf{Sampling aliases and amplitude response to compensation.} (a,b) Equivalent-height amplitude spectra of the observations, of the observations after subtraction of the jointly fitted donor waveform, and of that donor alone, on identical timestamps and nuisance projection, with 0--6 cm zooms below. Coloured vertical lines mark three off-centre candidate frequencies selected from the pure-injection response. (c,d) Observed amplitude at those frequencies as the compensation factor varies. Amplitude axes are linear and share scales within rows.}
\end{figure*}

\begin{figure*}[p]
\includegraphics[width=\textwidth]{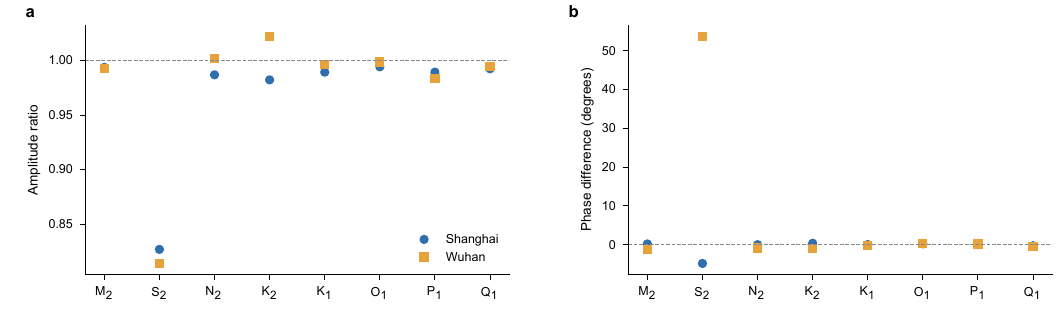}
\caption{\label{figureS7}\textbf{Radial-displacement validation of the loading convolution.} (a,b) Amplitude ratios (\textbf{a}) and phase differences (\textbf{b}) between radial displacements calculated with the loading Green's function and the FES2022 load-tide grid in the centre-of-mass frame, at Shanghai and Wuhan, for eight constituents. The S\textsubscript{2} difference is discussed in Supplemental Material,~\ref{sm:loading-validation}.}
\end{figure*}

\begin{table*}[p]
\caption{\label{tableS1}\textbf{Main and filtered validation analyses.} Both reductions use the same 54 observing segments, each with its own offset. They differ in selection, filtering and weighting.}
\includegraphics[width=\textwidth]{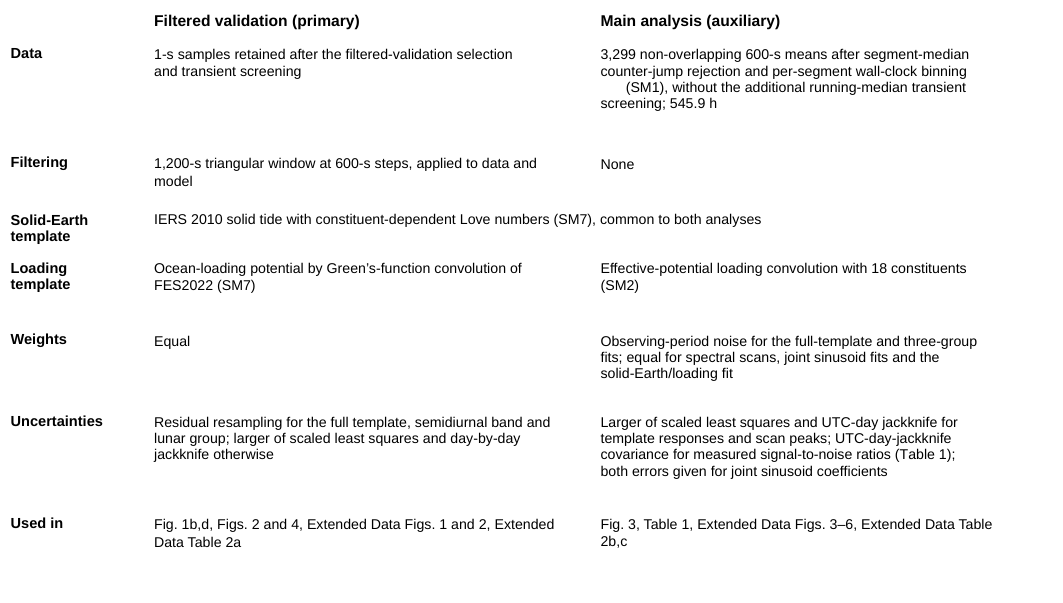}
\end{table*}

\begin{table*}[p]
\caption{\label{tableS2}\textbf{Campaign responses, joint coefficients and signal-to-noise ratios.} (a) Daily terms and campaign-specific tidal amplitudes. Amplitude and local time of maximum of the 24-h term, and amplitude of the 12-h term, fitted jointly with the full tidal model within each campaign. Days are local calendar days with data. Daily amplitudes are expressed in 10\textsuperscript{−18}. Their uncertainties are the larger of the scaled least-squares and day-by-day jackknife values; uncertainties in the time of maximum are jackknife estimates. Weak sinusoidal amplitudes are biased positive. The full-template amplitude \emph{A}, fitted without daily terms, and its residual-calibrated uncertainty are those in Fig. S1a. The last column gives the lunar amplitude fitted independently to each campaign with the filtered-validation estimator. (b) Free sinusoidal coefficients in the main analysis. Joint M\textsubscript{2} and S\textsubscript{2} cosine--sine coefficients, after controlling 54 segment constants and freely phased N\textsubscript{2} and K\textsubscript{2} terms. Coefficients and standard errors are in units of \(10^{-18}\). Phases refer to 6 September 2025, 16:05 UTC. Scaled least-squares errors use the residual variance; UTC-day jackknife errors delete each of the 43 observed dates and refit all coefficients. (c) Template-response signal-to-noise ratios. The first four rows use the non-overlapping full-template or three-group fits; their empirical errors are the larger of scaled-least-squares and UTC-day-jackknife errors. The final two rows use the equal-weight complete solid-Earth/loading fit with UTC-day-jackknife errors. The common-white-noise calculation retains each estimator and uses \(\alpha=5.4\times10^{-16}\sqrt{\mathrm{s}}\). The dimensionless factor \(\eta_i\) is defined in Supplemental Material,~\ref{sm:signal-to-noise}.}
\includegraphics[width=\textwidth]{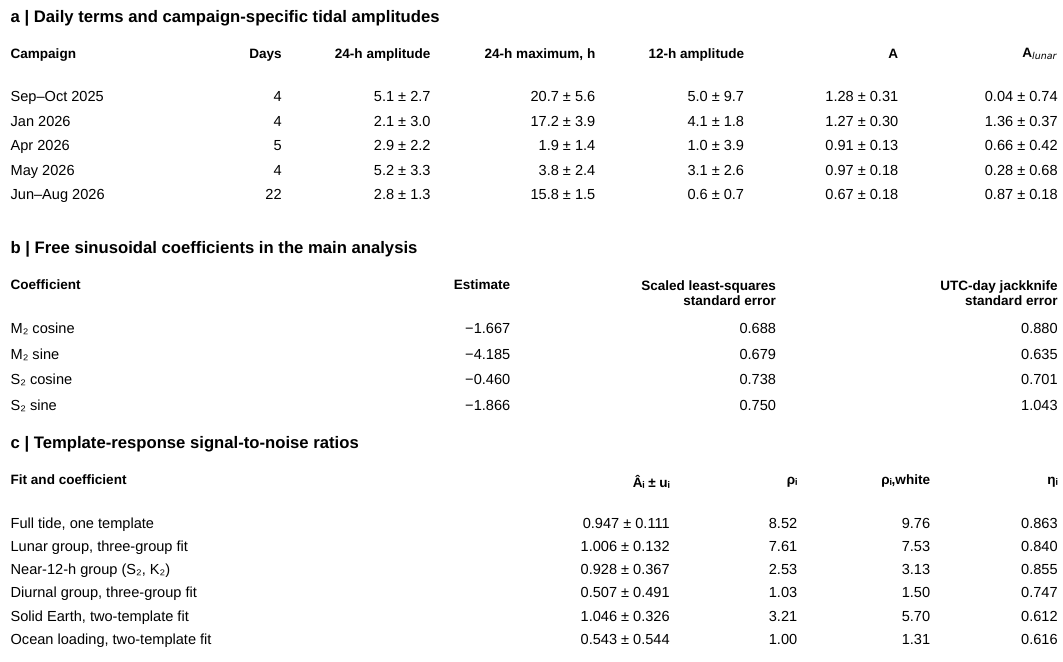}
\end{table*}

\clearpage
\section*{Supplemental References}
\begin{list}{}{\setlength{\leftmargin}{2.2em}\setlength{\labelwidth}{2em}\setlength{\itemsep}{1pt}\small}
\item[{[1]}] Jia, Z.-P. et al.~Improved systematic evaluation of a strontium optical clock with uncertainty below \(1\times10^{-18}\). \emph{Metrologia} \textbf{63}, 025002 (2026).
\item[{[2]}] Zhu, Q. et al.~Ytterbium lattice clock with systematic uncertainty of \(1.3\times10^{-18}\) and instability at the \(10^{-19}\) level. \emph{Metrologia} \textbf{63}, 065002 (2026).
\item[{[3]}] Hou, L. et al.~Yb/Sr optical clock comparison at \(2.2\times10^{-18}\) uncertainty over a 1350-km fiber link. Preprint at \url{https://arxiv.org/abs/2610.34250v1} (2026).
\item[{[4]}] Petit, G. \& Luzum, B. (eds) \emph{IERS Conventions (2010)}. IERS Technical Note \textbf{36} (2010). \url{https://iers-conventions.obspm.fr/content/tn36.pdf}
\item[{[5]}] Lyard, F. H., Allain, D. J., Cancet, M., Carrère, L. \& Picot, N. FES2014 global ocean tide atlas: design and performance. \emph{Ocean Sci.} \textbf{17}, 615--649 (2021).
\item[{[6]}] CNES. \emph{FES2022 (Finite Element Solution) Tidal model (Version 2024)} {[}Data set{]} (2024). \url{https://doi.org/10.24400/527896/A01-2024.004}
\item[{[7]}] Farrell, W. E. Deformation of the Earth by surface loads. \emph{Rev.~Geophys.} \textbf{10}, 761--797 (1972).
\item[{[8]}] Han, D. \& Wahr, J. The viscoelastic relaxation of a realistically stratified earth, and a further analysis of postglacial rebound. \emph{Geophys. J. Int.} \textbf{120}, 287--311 (1995).
\item[{[9]}] Sutterley, T., Velicogna, I., Mohajerani, Y. \& Lecomte, H. \emph{gravity-toolkit}, version 1.2.8. Zenodo \url{https://doi.org/10.5281/zenodo.23107010} (2026).
\item[{[10]}] Riley, W. J. \emph{Handbook of Frequency Stability Analysis}. NIST Special Publication \textbf{1065} (2008). \url{https://doi.org/10.6028/NIST.SP.1065}
\item[{[11]}] Birge, R. T. The calculation of errors by the method of least squares. \emph{Phys. Rev.} \textbf{40}, 207--227 (1932).
\item[{[12]}] Hersbach, H. et al.~The ERA5 global reanalysis. \emph{Q. J. R. Meteorol. Soc.} \textbf{146}, 1999--2049 (2020).
\end{list}

\begin{thebibliography}{99}
\bibitem{r3} Chou, C. W., Hume, D. B., Rosenband, T. \& Wineland, D. J. Optical clocks and relativity. \emph{Science} \textbf{329}, 1630--1633 (2010).
\bibitem{r10} Arnold, K. J. et al.~Lu\(^+\) optical frequency references with accuracy verified at the 19th digit. \emph{Nature} \textbf{658}, 60--64 (2026).
\bibitem{r5} McGrew, W. F. et al.~Atomic clock performance enabling geodesy below the centimetre level. \emph{Nature} \textbf{564}, 87--90 (2018).
\bibitem{r11} Bothwell, T. et al.~Resolving the gravitational redshift across a millimetre-scale atomic sample. \emph{Nature} \textbf{602}, 420--424 (2022).
\bibitem{r4} Lisdat, C. et al.~A clock network for geodesy and fundamental science. \emph{Nat. Commun.} \textbf{7}, 12443 (2016).
\bibitem{r12} Shen, Q. et al.~Free-space dissemination of time and frequency with \(10^{-19}\) instability over 113 km. \emph{Nature} \textbf{610}, 661--666 (2022).
\bibitem{r13} Caldwell, E. D. et al.~Quantum-limited optical time transfer for future geosynchronous links. \emph{Nature} \textbf{618}, 721--726 (2023).
\bibitem{r14} Gozzard, D. Clocks synchronized at the quantum limit. \emph{Nature} \textbf{618}, 680--681 (2023).
\bibitem{r15} Kómár, P. et al.~A quantum network of clocks. \emph{Nat. Phys.} \textbf{10}, 582--587 (2014).
\bibitem{r16} Nichol, B. C. et al.~An elementary quantum network of entangled optical atomic clocks. \emph{Nature} \textbf{609}, 689--694 (2022).
\bibitem{r17} Grotti, J. et al.~Geodesy and metrology with a transportable optical clock. \emph{Nat. Phys.} \textbf{14}, 437--441 (2018).
\bibitem{r18} Tapley, B. D., Bettadpur, S., Ries, J. C., Thompson, P. F. \& Watkins, M. M. GRACE measurements of mass variability in the Earth system. \emph{Science} \textbf{305}, 503--505 (2004).
\bibitem{r19} Rymer, H. \& Brown, G. Gravity changes as a precursor to volcanic eruption at Poás volcano, Costa Rica. \emph{Nature} \textbf{342}, 902--905 (1989).
\bibitem{r20} Lynett, P. et al.~Diverse tsunamigenesis triggered by the Hunga Tonga-Hunga Ha'apai eruption. \emph{Nature} \textbf{609}, 728--733 (2022).
\bibitem{r21} Rodell, M. et al.~Emerging trends in global freshwater availability. \emph{Nature} \textbf{557}, 651--659 (2018).
\bibitem{r6} Jia, Z.-P. et al.~Improved systematic evaluation of a strontium optical clock with uncertainty below \(1\times10^{-18}\). \emph{Metrologia} \textbf{63}, 025002 (2026).
\bibitem{r7} Zhu, Q. et al.~Ytterbium lattice clock with systematic uncertainty of \(1.3\times10^{-18}\) and instability at the \(10^{-19}\) level. \emph{Metrologia} \textbf{63}, 065002 (2026).
\bibitem{r22} Chen, F.-X. et al.~\(10^{-21}\)-Level optical frequency dissemination over 2067 km of noise-loaded field-deployed fiber network. \emph{Light Sci. Appl.} \textbf{15}, 276 (2026).
\bibitem{r23} Hou, L. et al.~Yb/Sr optical clock comparison at \(2.2\times10^{-18}\) uncertainty over a 1350-km fiber link. Preprint at \url{https://arxiv.org/abs/2610.34250v1} (2026).
\bibitem{r1} Petit, G. \& Luzum, B. (eds) \emph{IERS Conventions (2010)}. IERS Technical Note \textbf{36} (2010). \url{https://iers-conventions.obspm.fr/content/tn36.pdf}
\bibitem{r24} Lyard, F. H., Allain, D. J., Cancet, M., Carrère, L. \& Picot, N. FES2014 global ocean tide atlas: design and performance. \emph{Ocean Sci.} \textbf{17}, 615--649 (2021).
\bibitem{r25} CNES. \emph{FES2022 (Finite Element Solution) Tidal model (Version 2024)} {[}Data set{]} (2024). \url{https://doi.org/10.24400/527896/A01-2024.004}
\bibitem{r2} Farrell, W. E. Deformation of the Earth by surface loads. \emph{Rev.~Geophys.} \textbf{10}, 761--797 (1972).
\bibitem{SM} See the Supplemental Material appended to this manuscript for measurement and record construction, physical predictions, response fits and uncertainties, spectral and constituent separation, the signal-to-noise calculation, daily-term and environmental checks, the filtered validation analysis, separation and sampling diagnostics, Figs. S1--S7 and Tables S1 and S2.
\bibitem{r8} Dimarcq, N. et al.~Roadmap towards the redefinition of the second. \emph{Metrologia} \textbf{61}, 012001 (2024).
\bibitem{r9} Wu, H., Müller, J. \& Lämmerzahl, C. Clock networks for height system unification: a simulation study. \emph{Geophys. J. Int.} \textbf{216}, 1594--1607 (2019).
\bibitem{r26} Stahl, K. et al.~Optical clock network tracks tidal variations of the gravity potential. Preprint at \url{https://arxiv.org/abs/2610.39635v1} (2026).
\end{thebibliography}
\end{document}